# Tuning Breakdown-to-Coercive Field Ratio in Ultra-Thin Al$_{1-x}$Sc$_x$N Films via Reactive Nitrogen Atmosphere

Yinuo Zhang, Walter J. Smith, Giovanni Esteves, *Eric A. Stach\*, Thomas E. Beechem,\* and Roy H. Olsson III\*

*Abstract*—Al$_{1-x}$Sc$_x$N has attracted significant interest due to its large remnant polarization and low processing temperature when compared to other ferroelectric material systems. However, device dielectric failure before ferroelectric switching remains a critical limitation for AlScN-based memory devices. With the continuing trend toward device miniaturization, expanding the operating window is essential for next-generation memory development. In this work, we optimized the breakdown field ($E_{BD}$) and coercive field ($E_C$) in ultra-thin Al$_{1-x}$Sc$_x$N films by controlling defect density via adjustment of nitrogen gas flow during sputter deposition. The characteristic breakdown field, $E_{BD}$, was evaluated using Weibull statistics, yielding optimal characteristic breakdown fields of 12.47 MV/cm ($E_{BD}^+$) and –12.63 MV/cm ($E_{BD}^-$) for samples deposited under 27.5 sccm N$_2$ flow. The minimum $E_C$ was achieved at a nitrogen flow of 25 sccm and increased for higher gas flows, a trend that is opposite to previous reports in much thicker films. The highest $E_{BD}/E_C$ ratio of 2.25 occurred at 27.5 sccm, effectively expanding the operational window. Using a combination of X-ray diffraction and photoluminescence spectroscopy to study changes in crystal orientation and defects, device performance can be tuned by controlling the point defect concentration in the ultra-thin film via adjusting the sputtering N$_2$ process gas flow rate.

I. Introduction

Ferroelectric aluminum scandium nitride (Al$_{1-x}$Sc$_x$N) has emerged as a promising material for next-generation non-volatile memories, owing to its high spontaneous polarization, scalability, and compatibility with complementary metal-oxide-semiconductor (CMOS) back-end-of-line (BEOL) processes.[1-5] These attributes make Al$_{1-x}$Sc$_x$N particularly attractive for integration into ultra-scaled ferroelectric devices. As the demand for minimizing device dimensions intensifies, maintaining and improving robust ferroelectric performance in ultra-thin Al$_{1-x}$Sc$_x$N films with thickness < 20 nm has become increasingly critical.[6-8] Various approaches, such as interfacial engineering, impurity doping, and defect density control, have been explored to improve and optimize ferroelectricity at reduced thicknesses.[9-12] Nitrogen gas flow is a critical processing parameter that directly influences film stoichiometry and crystallinity and is known to influence the ferroelectric properties in thicker Al$_{1-x}$Sc$_x$N films.[13-15] Despite this, the role of nitrogen gas (N$_2$) flow during co-sputtering deposition on the ferroelectric behavior of ultra-thin Al$_{1-x}$Sc$_x$N has not been investigated.

For example, *Yang et al.* demonstrated that the crystalline quality of 2 μm thick *c*-axis oriented AlScN deposited by reactive pulsed Direct Current (DC) magnetron sputtering could be optimized by changing N$_2$/Ar sputtering atmosphere, which also affected the film electrical properties such as resistivity.[16] The electromechanical properties are also influenced by reactive gas concentration in thicker Al$_{1-x}$Sc$_x$N.[17] Even with this clear dependency between reactive gas environment and the electrical response in thicker Al$_{1-x}$Sc$_x$N, its impact on key device performance metrics such as coercive field ($E_C$), breakdown field ($E_{BD}$), and their ratio ($E_{BD}/E_C$) in the ultra-thin (~10 nm) thickness regime characteristic of scaled memory devices remains unclear. Moreover, a mechanistic description linking the N$_2$ flow rate to material changes in the AlScN has yet to be established. Here, we examine the impact of N$_2$ gas flow environment on the electrical characteristics of

This work is supported in part by Army/ARL via the Collaborative for Hierarchical Agile and Responsive Materials (CHARM) under cooperative agreement W911NF-19-2-0119. This work was performed in part at the Singh Center for Nanotechnology at the University of Pennsylvania, a member of the National Nanotechnology Coordinated Infrastructure (NNCI) network, which is supported by the National Science Foundation (Grant NNCI-2025608). Additional support for the X-ray diffraction facility at the University of Pennsylvania was provided by the NSF via the Materials Research Science and Engineering Center (MRSEC) (DMR-2309043). This research was primarily/partially supported by NSF through the University of Pennsylvania Materials Research Science and Engineering Center (MRSEC) (DMR-2309043). W.S. and T.B. acknowledge funding from Sandia National Laboratories University Partnerships program administered through Laboratory Directed Research and Development (LDRD). Yinuo Zhang and Roy H. Olsson are with the Electrical Systems and Engineering Department, University of Pennsylvania, Philadelphia, PA 19104 USA. (e-mail: rolsson@seas.upenn.edu).

Eric. A Stach is with the Material Science and Engineering Department, University of Pennsylvania, Philadelphia, PA 19104 USA. (e-mail: stach@seas.upenn.edu).

Walter J. Smith and Thomas J. Beechem are with the School of Mechanical Engineering, Purdue University and Birck Nanotechnology Center, West Lafayette, IN 47907 USA, (e-mail: tbeechem@purdue.edu).

Giovanni Esteves is with the Microsystems Engineering, Science and Applications, Sandia National Laboratories, Albuquerque, New Mexico, USA.

ultra-thin Al$_{1-x}$Sc$_x$N and assess the material changes causing the observed changes in device characteristics.

The reactive gas environment can influence the point defect concentration (*e.g.,* nitrogen (N) or metal (M) vacancies) in the film, which impacts device performance. Vacancies in AlN-films, for example, have been reported to modulate the bandgap and conductivity, increase leakage current, and cause ferroelectric switching fatigue.[18-20] With respect to Al$_{1-x}$Sc$_x$N in particular, a recent study of a 100 nm thick film suggested that a reduction in N-vacancies induced by a N-rich TiN bottom electrode improved ferroelectric performance.[21] From this evidence, we hypothesize that N and metal (i.e., Al or Sc) vacancy density in ultra-thin (~10 nm) Al$_{1-x}$Sc$_x$N films can be tuned by varying the N$_2$ partial pressure while removing Ar gas during the co-sputtering process, leading to corresponding improvement in the electrical properties of the ferroelectric Al$_{1-x}$Sc$_x$N material.

To test this hypothesis, E$_C$, E$_{BD}$ and their ratio (E$_{BD}$/E$_C$) were measured as a function of N$_2$ gas flow during synthesis to assess the degree to which this process parameter might affect the switching performance, operational window and device reliability. These parameters are particularly important for practical device design in ultra-thin film systems, where increased leakage current and higher switching barriers present significant challenges.[22-24]

Beyond assessing device characteristics, material alterations induced by changes in the reactive gas environment were also examined via photoluminescent studies capable of tracking changes in defects. Using this approach, moderate N$_2$ flow rates were found to improve the E$_{BD}$ and E$_{BD}$/E$_C$ ratio. This performance improvement correlated with an increase in the PL signature associated with N interstitials defects suggesting a lowering of N- vacancy concentration correlated with the higher performance. However, as the N$_2$ flow rate was further increased, the overall PL intensity increased, suggesting an increase in overall defect concentration that correlates with measured reductions in the E$_{BD}$ and E$_{BD}$/E$_C$.

The paper is organized as follows. In Section II we present the experimental details, including the film growth conditions, device fabrication process, material and electrical characterization procedures. In Section III, we present the results and correlate the film growth orientation, device electrical properties, and PL results to elucidate the impact of defects on ferroelectric failure and switching performance.

## II. Experimental Details

### A. Film Growth, Device Fabrication and Characterization

Ultra-thin Al/Al$_{0.68}$Sc$_{0.32}$N/Al (50 nm/~10 nm/50 nm) film stacks were deposited on 4 inch *c*-axis oriented sapphire wafers using the target powers, temperature, time and target-to-substrate distance reported in Zhang et al.[9] and summarized in Table 1. The N$_2$ gas flow was held constant during each individual deposition. The entire Al/Al$_{0.68}$Sc$_{0.32}$N/Al was deposited without vacuum break to prevent surface oxidation of the Al$_{0.68}$Sc$_{0.32}$N and Al bottom electrode as reported by Zheng et al.[7] The film stacks for photoluminescence analysis are sapphire/Al(50 nm)/ Al$_{0.68}$Sc$_{0.32}$N (200nm)/AlN(20nm). The 200 nm Al$_{0.68}$Sc$_{0.32}$N films were grown with the same target powers but each sample was grown under a different N$_2$ gas flow from 20 sccm to 40 sccm (20, 25, 30, 35, 40 sccm) followed by a 20 nm AlN deposition to prevent oxidation of Al$_{0.68}$Sc$_{0.32}$N. Ferroelectric capacitors (FeCaps) were fabricated using the process reported in Zhang et al.[9]

Table 1. Deposition parameters of ultra-thin Al$_{0.68}$Sc$_{0.32}$N

| Parameter | Value/Condition |
|---|---|
| Substrate | Sapphire |
| Target power density (W/cm²) | Al: 12.7 W/cm² <br> Sc: 8.32 W/cm² |
| Temperature (°C) | 350 |
| Deposition Time (s) | 42.5 |
| Base Pressure (mbar) | ~ 5×10$^{-8}$ |
| Target-to-substrate distance (mm) | 103.3 |
| N$_2$ gas flow (sccm) | 22.5, 25, 27.5, 30, 35, 40 |

### B. X-Ray Diffraction

X-ray diffraction (XRD) and X-ray reflectivity (XRR) were used to evaluate AlScN film thickness and texture using a PANalytical X'Pert³ MRD diffractometer equipped with a monochromatic beam and PIXcel detector. $\theta$–$2\theta$ scans were acquired to determine the out-of-plane orientation and lattice spacing of the AlScN and Al layers. Rocking curve ($\omega$-scan) measurements of the AlScN (0002) reflection were conducted to assess crystallographic texture, where narrower full width at half maximum (FWHM) values indicate stronger alignment. The (0002) rocking curve peaks were fit to extract FWHM values for sample comparison, while XRR data were modeled in GSAS-II to fit Kiessig fringes and obtain film thickness.

### C. Electrical Measurement

The breakdown field values were collected by performing 25 kHz (pulse width = 2×10$^{-5}$ s) monophasic triangular pulses on 20 spatially distributed capacitors across the wafer. The positive electric breakdown (E$_b^+$) and negative electric breakdown fields (E$_b^-$) were obtained by applying pulses with opposite electric field polarizations. The coercive fields were obtained by 25 kHz biphasic triangular Positive-Up-Negative-Down (PUND) pulses,[23] on 5 different capacitors spatially distributed across the wafer. All the electrical tests were conducted using a Keithley 4200-SCS system.

*D. Photoluminescence analysis*

The evolution of defects were studied via photoluminescence spectroscopy (PL) on films with varying $N_2$ flow during film growth. PL was collected in air on a WITec Alpha300R confocal microscope using a 405 nm laser at 0.10 mW across a diffraction limit spot size of ~450 nm. Scattered light was collected using a UHTS VIS 600 spectrometer with a spectral resolution of less than 0.01 eV. Fitting of the spectra and assignment of features was performed using methods analogous to that employed in [20].

### III. RESULTS AND DISCUSSION

*A. X-Ray Diffraction and X-Ray Reflectometry Analysis*

The crystallographic orientation of the films was studied using X-ray diffraction (XRD). As shown in **Figure 1a**, a 2θ scan from 30° to 40° was performed on Al/Al$_{0.68}$Sc$_{0.32}$N/Al film stacks. The (0002) out-of-plane diffraction peak of Al$_{0.68}$Sc$_{0.32}$N was clearly identified at approximately 35.5°, while a highly textured Al(111) peak was observed near 38.5°. To further evaluate the c-axis orientation, rocking curve scans were conducted for each sample, as shown in Figure 1b. The full-width at half-maximum (FWHM) values were extracted from these curves with their dependence on reactive $N_2$ flow rate is exhibited in Figure 1c. Overall, the FWHM values increase with increasing $N_2$ flow rate, suggesting a lower quality of *c*-axis oriented AlScN except the film grown under 35 sccm $N_2$ which shows a significant increase in the FWHM and corresponds to the device with the poorest dielectric breakdown performance, as will be discussed in Section B. In contrast, the film deposited under the lowest $N_2$ flow condition exhibited the narrowest FWHM, indicating a better c-axis alignment of the Al$_{0.68}$Sc$_{0.32}$N (0002).

To gain information on film thickness for property analysis, X-ray Reflectometry (XRR) measurements were conducted on each layer stack. The thickness of Al$_{0.68}$Sc$_{0.32}$N is summarized in **Table 2.** The thickness decreases with increasing $N_2$ gas flow, even though all films were deposited for the same duration, an effect attributed to increased scattering at higher flow rates. The thickness of the ultra-thin Al$_{0.68}$Sc$_{0.32}$N films ranged from 9.6 nm to 11.7 nm.

*B. Breakdown and Failure Analysis*

Subsequent to thin film deposition and structural characterization, the ferroelectric performance and device reliability were studied. Electrical measurements were performed on ferroelectric capacitors with a 4 μm diameter. A cross-sectional diagram of the capacitor structure is depicted in **Figure 2**.

First, the impact of $N_2$ flow on breakdown field was investigated. The Weibull distribution of the FeCap breakdown fields for the films deposited under different $N_2$ flows is plotted in **Figure 3a** and **Figure 3b.** The Weibull probability distribution is commonly used to describe the statistical distribution of electron trap density that leads to dielectric failure [25] and is widely employed to assess the lifetime of electronic devices and to predict the reliability of manufactured products.[26-28] The relationship between the failure probability and experimental breakdown field data points is described by equation (1),

$$F = 1 - \exp\left(-\left(\frac{E_b}{E_{BD}}\right)^\beta\right) \quad (1)$$

where $E_b$ is the experimental breakdown field data points. Each data point was obtained by applying a unipolar, 25 kHz

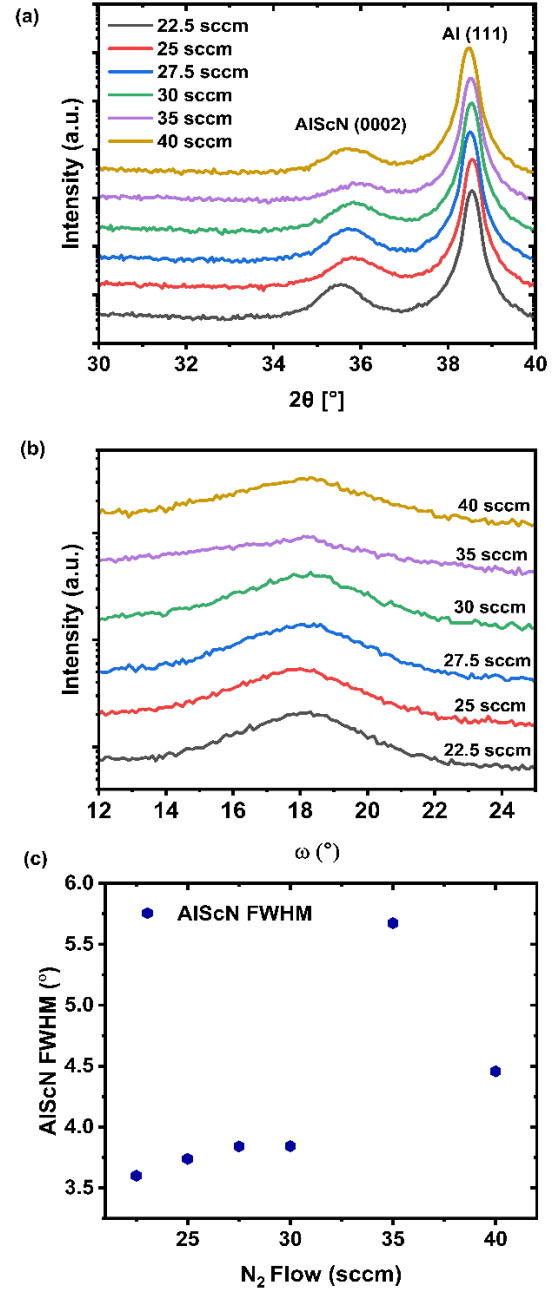

**Figure 1.** XRD measurement of thin film stacks. (a). 2θ scan of Al/Al$_{0.68}$Sc$_{0.32}$N/Al from 30° to 40° with (0002) Al$_{0.68}$Sc$_{0.32}$N and (111) Al labeled. (b). Rocking curve scan for the (0002) Al$_{0.68}$Sc$_{0.32}$N films deposited under different $N_2$ flow. (c) FWHM obtained by Pearson VII for ultra-thin Al$_{0.68}$Sc$_{0.32}$N films by varying the reactive $N_2$ gas flow.

triangular pulse with a maximum voltage equal to +40 V. The breakdown failure of the device was evaluated by both applying positive and negative monopolar triangular waveforms. Twenty capacitors spatially distributed across the full area of a wafer piece were selected randomly as the breakdown test subjects. $E_{BD}$ is the breakdown field where the failure probability is 63% (also called characteristic breakdown field). [28, 29] In this work, we used $E_{BD}$ to compare the breakdown properties of each device and calculated $E_{BD}/E_C$ to evaluate the window of operation and endurance potential.[30, 31] β is the shape factor of the Weibull distribution plot, which describes the uniformity the breakdown field.[25, 32]

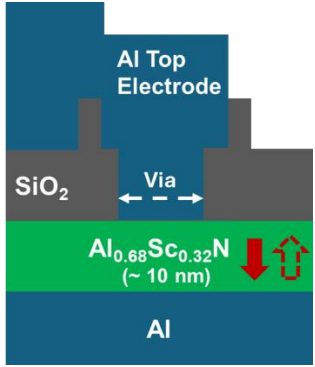

**Figure 2.** Cross-section schematic of FeCap structure. The dashed arrow in the Al$_{0.68}$Sc$_{0.32}$N layer represents the ferroelectric dipole direction after deposition while the solid red arrow represents the direction of the dipole after switching.

Characteristic breakdown values ($E_{BD}^+$ and $E_{BD}^-$) as a function of N$_2$ flow are plotted in **Figure 3c.** The FeCaps fabricated from the ultra-thin film deposited under 27.5 sccm N$_2$ exhibit the highest $E_{BD}$ under both pulse polarities (12.47 MV/cm and -12.63 MV/cm). Under positive triangular wave pulses, $E_{BD}$ decreased with further increases in N$_2$ flow. For instance, at 40 sccm, $E_{BD}$ dropped to 9.9 MV/cm, the lowest value observed among all tested devices. While under negative pulses, a pronounced reduction could also be observed as N$_2$ flow increased beyond 27.5 sccm, but the absolute $E_{BD}^-$ values starts to increase again after reaching a minimum for the 35 sccm N$_2$ sample.

To evaluate the breakdown uniformity, the shape factor, β, was extracted from each Weibull Distribution and is summarized in **Table 2**. All devices under positive and negative pulses exhibit shape factors greater than 1, suggesting that the failure probability increases with higher electric field ($E_b$), consistent with a wear-out breakdown mechanism caused by high-voltage pulse stress.[33, 34] Furthermore, the FeCaps fabricated from the ultra-thin AlScN film grown under 27.5 sccm N$_2$ have the highest shape factor (β$^+$ = 35.3, β$^-$ = 16.50), producing the most steep Weibull curve and indicating a tight breakdown field distribution with excellent device-to-device consistency. In contrast, the lowest β$^+$ and β$^-$ generated by the FeCaps were observed in the ferroelectric Al$_{0.68}$Sc$_{0.32}$N layer deposited at 40 sccm N$_2$ and 35 sccm N$_2$ flow respectively, indicating a broad breakdown distribution and a wider range of breakdown fields, especially for applied voltage pulses with positive

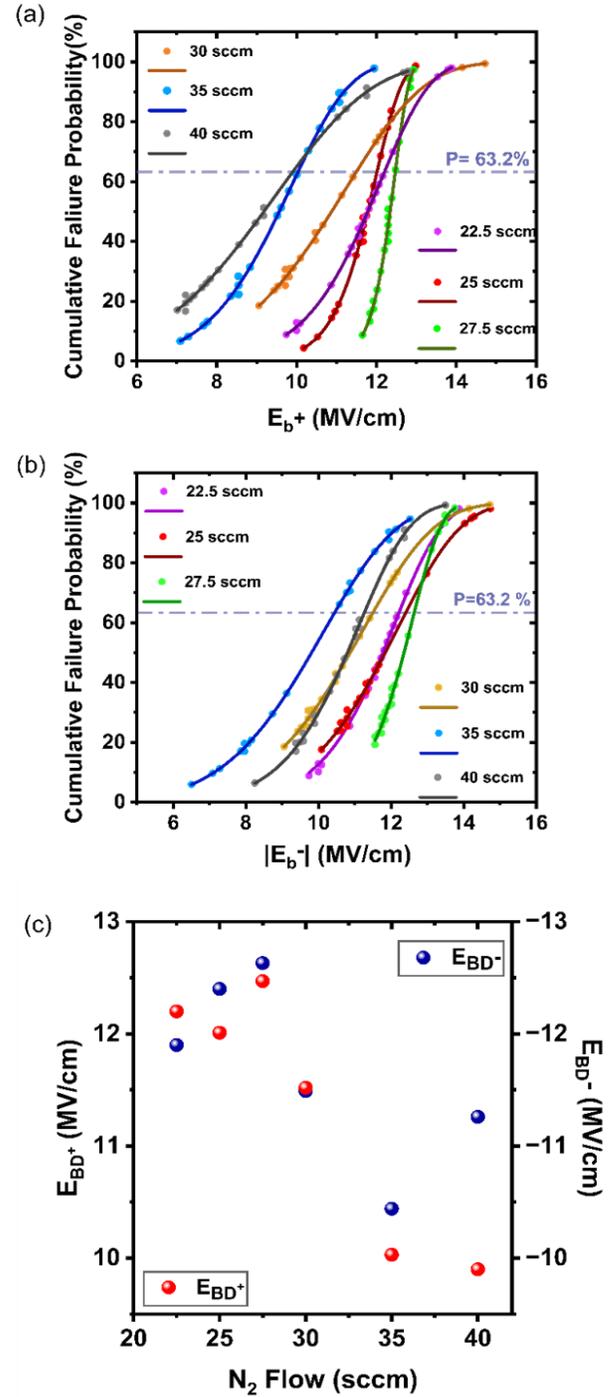

**Figure 3.** Breakdown analysis of Al$_{0.68}$Sc$_{0.32}$N FeCaps (a) Weibull distribution of FeCap breakdown when applying positive triangular pulses. (b) Weibull distribution of FeCap breakdown when applying negative triangular pulses. (c) $E_{BD}^+$ and $E_{BD}^-$ versus N$_2$ gas flow

polarity. The $E_{BD}$ and shape factors of the devices made from different sputtering atmospheres are summarized in **Table 2**. Overall, the 27.5 sccm N$_2$ flow condition yielded both the highest breakdown field and the most concentrated failure distribution. Moreover, devices grown under higher N$_2$ flow rates (i.e., 35 sccm and 40 sccm N$_2$), which generated higher XRD-derived FWHM values and thus less c-axis oriented AlScN (see Figure 1b and 1c) exhibited poorer breakdown performance. Simply stated, greater disorder in the films correlated with reduced breakdown performance and uniformity in device response, as exemplified by the lower β for these films (see **Table 2**). This behavior suggests increased defect activity, which is supported by the PL-measurements discussed in Section D.

**Table 2.** Summary of the electrical performance of FeCaps deposited under different N$_2$ flows.

| N$_2$Flow (sccm) | 22.5 | 25 | 27.5 | 30 | 35 | 40 |
|---|---|---|---|---|---|---|
| Thickness (nm) | 11.7 | 11.4 | 10.9 | 10.6 | 10.4 | 9.6 |
| E$_{BD+}$ (MV/cm) | 12.2 | 12.0 | 12.5 | 11.5 | 10.0 | 9.9 |
| E$_{BD-}$ (MV/cm) | -11.9 | -12.4 | -12.6 | -11.5 | -10.4 | -11.3 |
| β+ (E$_{BD+}$) | 10.6 | 18.7 | 35.3 | 6.6 | 7.7 | 4.9 |
| β- (E$_{BD-}$) | 8.2 | 8.0 | 16.5 | 6.6 | 5.9 | 8.7 |
| E$_C^-$ (MV/cm) | -5.7 | -5.5 | -5.7 | -5.7 | -5.8 | -6.0 |
| E$_{BD-}$/E$_C^-$ | 2.1 | 2.3 | 2.2 | 2.0 | 1.8 | 1.9 |

## C. Coercive Field Measurements and Device Operating Window Evaluation

The E$_C$ of films deposited with different N$_2$ flow rates was investigated via triangular Positive-Up-Negative-Down (PUND) tests. A 25 kHz triangular Current density-Electric field (JE) JE-PUND analysis was employed to compensate leakage current and isolate the ferroelectric switching response.[9, 35] To minimize selection bias, five FeCaps with spatially distributed locations on each wafer piece were selected for E$_C$ measurements. The average values and corresponding error bars are presented in **Figure 4**.

The positive coercive field could not be clearly identified due to measurement limitations and increased leakage during the Nitrogen-polar to Metal-polar polarization switching of the ultra-thin films. Therefore, only the negative coercive field was considered for comparative analysis across different N$_2$ flow conditions. As shown in **Figure 4**, although the film grown under 22.5 sccm N$_2$ exhibited a slightly higher E$_C$, a clear increasing trend in coercive field was observed starting from 25 sccm N$_2$ flow. This is counter to the trend observed in thicker films (> 50 nm) where an increased N$_2$ flow led to a more tensile film stress and a lower coercive field. [3, 17].

To better evaluate the device's operational window, the absolute value of E$_{BD}$/E$_C$ with N$_2$ flow during growth is plotted in **Figure 5**, where the average E$_C$ was selected to calculate E$_{BD}$/E$_C$. In the ultra-thin film system, a maximum E$_{BD}^-$/E$_C^-$ = 2.25 was obtained when the N$_2$ flow was 25 sccm. The best operation window occurs under the same conditions as that which leads to the greatest endurance potential. At lower N$_2$ flows (22.5 sccm), the ratio was slightly reduced. In contrast, a sharp decline in E$_{BD}^-$/E$_C^-$ value was observed when the N$_2$ concentration exceeded 27.5 sccm, which contradicts the trend reported by *Yang et al*. [21] measured in much thicker films. At 35 sccm N$_2$ flow, the ratio dropped to 1.80, the lowest value recorded in this study. The higher XRD FWHM value of *c*-axis oriented AlScN film grown under 35 sccm N$_2$ flow corresponds to the lowest E$_{BD-}$, suggesting that the quality of the film plays a role in inducing the dielectric failure. The $E_{BD}/E_C$ values each of device have been summarized in **Table 2**.

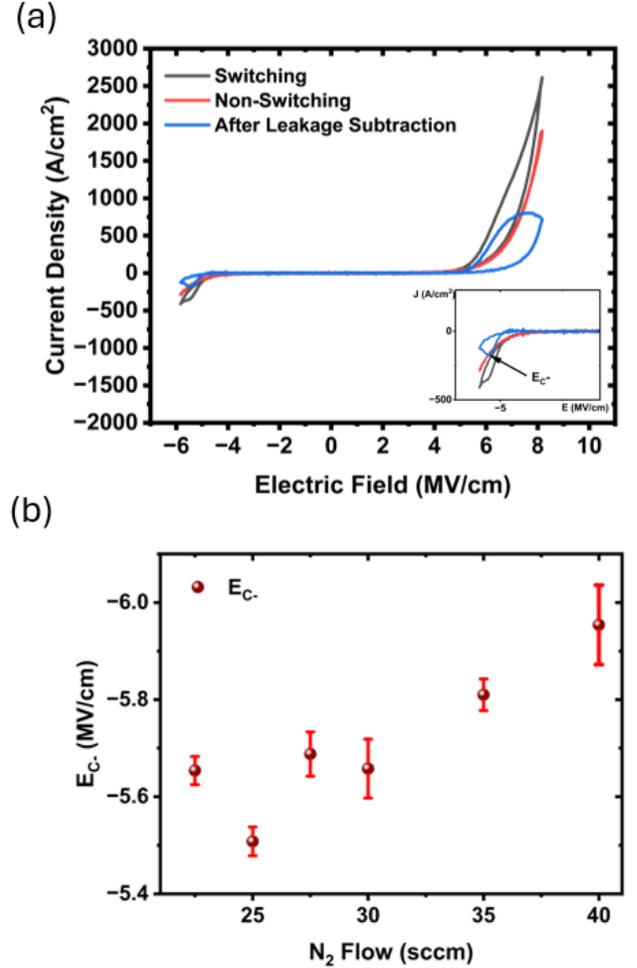

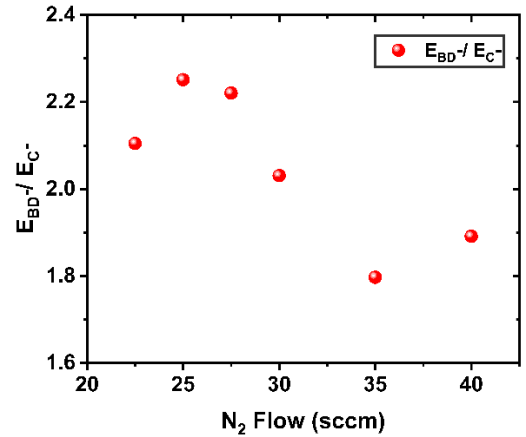

**Figure 4.** (a) JE-PUND measurment of 25 sccm N$_2$ flow rate AlScN showing switching, non-switching and JE loop after leakage subtraction; Insect is the zoom-in negative JE loop of E$_C$ location; (b) Negative E$_C$ versus N$_2$ gas flow

**Figure 5.** E$_{BD}^-$/E$_C^-$ versus N$_2$ flow

### D. Defect Analysis Using Photoluminescence

To understand the underlying mechanism for the $E_{BD}$ and $E_C$ trends as a function of $N_2$ flow during film growth, photoluminescence spectroscopy of the Al$_{0.68}$Sc$_{0.32}$N films was conducted and the spectra were analyzed via curve fitting. Photoluminescence is a three step-process, requiring absorption of incident photons, scattering through phonons or other means, and subsequent radiative recombination, which is measured as transitions that occur between the conduction band, valence band, and defects. Utilization of sub-bandgap excitation preferentially samples from defects as it prohibits band-to-band absorption for the visible laser used for excitation here.

Changes in the PL-spectra with increasing $N_2$ flow are provided in **Figure 6a** along with a representative fit used to assess the transitions giving rise to the signal (**Figure 6b**). Fitting was accomplished using a summation of four Gaussian functions revealing transitions having energies of 1.84 ± 0.10 eV, 1.95 ± 0.07 eV, 2.45 ± 0.07 eV, and 2.65 ± 0.03 eV. The linewidths (i.e. full-width at half-maximum, FWHM) for these transitions were 0.22 ± 0.05 eV, 0.49 ± 0.09 eV, 0.33 ± 0.06 eV, and 0.17 ± 0.03 eV.

As $N_2$ flow increases, the intensity of the 1.95 eV peak increases drastically while the intensity of the 2.45 eV and 2.65 eV peaks remain relatively constant, as shown in **Figure 7**. The intensity of the 1.84 eV, meanwhile, exhibits a non-monotonic dependence transitioning from a maximum between 20 sccm and 30 sccm, before subsequently decreasing. This non-monotonic behavior in the 1.84 eV feature is qualitatively similar to that observed both in the dependence of breakdown field (**Figure 3c**) and operation window (**Figure 5**) suggesting a link between this defect and overall performance.

To gain insight into the origin of this signal, we note that the formation energy, and thus the concentration of defects changes as the growth becomes increasingly N-rich. For example, increasing N during growth will decrease the number of N vacancies ($V_N$) and O substitutional defects ($O_N$). [36, 37] On the other hand, it will increase the probability of forming nitrogen interstitials, the number of metal vacancy defects ($V_{Al}$, $V_{Sc}$) and associated complexes consisting of these vacancies with any residual oxygen (i.e., oxygen defect complexes). [38] Recognizing this, changes in PL intensity with $N_2$ flow likely are linked to variations in these particular defects even as definitive assignment is quite difficult because the exact positions of many defects within the bandgap of AlScN remains unknown.

With this perspective, we can speculate that the constant increase in intensity for the transition at 1.95 with $N_2$-flow is likely caused by an increase in those defects whose formation energy is most affected, namely the nitrogen interstitial defects ($N_i$) and/or metal vacancies. [19, 36] This type of $N_i$ point defect is likely the key factor responsible for the rising $E_C^-$ with the increasing process $N_2$ gas flow (Figure 7a and 7b). Interestingly, the $E_C^-$ change of the ultra-thin Al$_{1-x}$Sc$_x$N film generally follows a linear trend with the energy transit intensity of 1.95 eV as shown in **Figure 7b**, indicating the correlation of defect density change tendency and ferroelectric switching energy barrier in the ultra-thin film system. Point defects are regarded as pinning sites during the domain wall motion in ferroelectric material systems. [39, 40] An increase in defects would, therefore, be expected to increase the coercive fields. However, their exact impact on ultra-thin AlScN film ferroelectric switching barrier remains unknown.

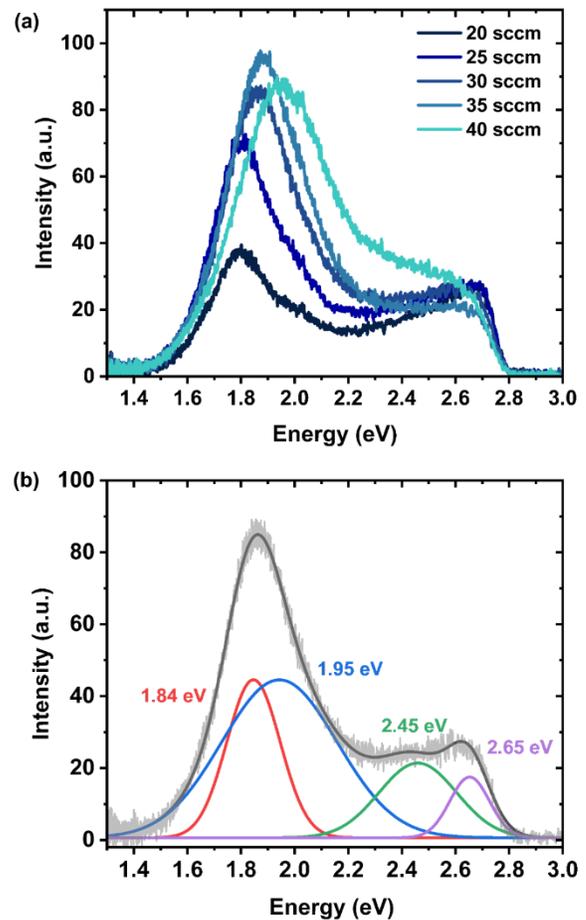

**Figure 6.** Photoluminescence spectrum of Al$_{0.68}$Sc$_{0.32}$N. (a) Signature peak intensity change vs. increasing $N_2$ flow rate. (b) Raw data of Al$_{0.68}$Sc$_{0.32}$N at 30 sccm $N_2$ flow with accompanying spectral fit comprised of four Gaussian line shapes.

The lower energy mode near 1.84 eV meanwhile is consistent with the presence of defect complexes composed of a metal vacancy interacting with residual oxygen (i.e., an O-complex).[38, 41] This deduction is based on knowledge that there is likely a small—but non-zero—concentration of oxygen in the films. This could be attributed to the replacement tendency of N by O during the film deposition.[42, 43] Furthermore, at low concentrations, even at low base pressures (< 5×10$^{-8}$ mbar), oxygen defects are known to almost always

"find" metal vacancies to pair with in AlN. [19] The non-monotonic response with N$_2$-flow suggests that the increasing amount of metal vacancies form O-complexes until the residual oxygen is fully utilized.[41] Regardless, the presence of either of these defects is associated with a reduction in N-vacancies.

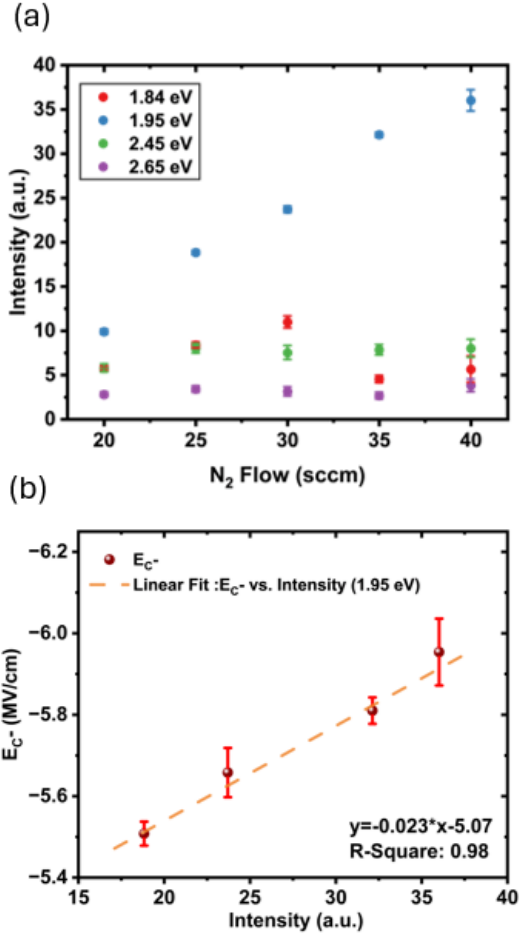

**Figure 7.** Energy transition correlates with E$_C^-$ change (a) Intensity of the different transitions in the Al$_{0.68}$Sc$_{0.32}$N films. The intensity of the peak at 1.95 eV increases steadily as N$_2$ flow increases while the intensity of the peak at 1.84 eV is a maximum at 30 sccm N$_2$. (b) E$_C^-$ vs. Energy Transition Intensity (1.95 eV) with a linear fitting.

The PL data, therefore, taken together with the E$_C$ and E$_{BD}$ data, suggests that a higher N$_2$ flow during film growth may decrease the number of N vacancies, resulting in an increase in E$_C$ with increasing N$_2$ flow. At higher N$_2$ flows, above 27.5 sccm, the density of other defects (at 1.84 eV) increases, which reduces the breakdown field. For example, even though having similar FWHM values, the film grown under 30 sccm N$_2$ has lower breakdown strength than the film grown under 27.5 sccm because of the higher concentration of the defect at 1.84 eV. Taken in aggregate, the PL highlights how the defect environments is tightly intertwined with the ultimate performance of the device.

## IV. CONCLUSION

In summary, we tuned the ferroelectric performance of ultra-thin AlScN capacitors by manipulating the processing N$_2$ gas flow to adjust the defect concentration. The breakdown field strength reached its maximum for an N$_2$ gas flow of 27.5 sccm. AlScN films with degraded c-axis texture indicated by higher XRD FWHM also exhibited lower dielectric breakdown. Interestingly, the coercive field increased with the increase in N$_2$ gas flow, which follows a opposite trend when compared to prior publications in thicker AlScN films. The E$_{BD^-}$/E$_{C^-}$ was optimized to 2.25 under a 25 sccm N$_2$ flow. To better understanding how the point defects impact the ferroelectricity and dielectric failure, photoluminescence spectroscopy was performed on a series of thicker films sputtered under the same conditions. Photoluminescence analysis indicates that under N-rich conditions N$_i$ related-defect energy transitions become more pronounced, exhibiting the same increasing trend of E$_{C^-}$ with N$_2$ gas flow. Even though exhibiting similar FWHM, the appearance of O-complexes leads to a lower breakdown strength at 30 sccm N$_2$ process gas flow when comparing to the film grown under 27.5 sccm N$_2$ flow. The degree of *c*-axis orientation plus the point defects density contributes to the observed reduction in breakdown strength as well as the variation of E$_{BD}$/E$_C$. These findings offer new understanding of the relationship between film texture, point defects and the reliability of AlScN-based ferroelectric memory devices.

## ACKNOWLEDGEMENT

Sandia National Laboratories is a multi-mission laboratory managed and operated by National Technology & Engineering Solutions of Sandia, LLC (NTESS), a wholly owned subsidiary of Honeywell International Inc., for the U.S. Department of Energy's National Nuclear Security Administration (DOE/NNSA) under contract DE-NA0003525. This written work is authored by an employee of NTESS. The employee, not NTESS, owns the right, title and interest in and to the written work and is responsible for its contents. Any subjective views or opinions that might be expressed in the written work do not necessarily represent the views of the U.S. Government. The publisher acknowledges that the U.S. Government retains a non-exclusive, paid-up, irrevocable, world-wide license to publish or reproduce the published form of this written work or allow others to do so, for U.S. Government purposes. The DOE will provide public access to results of federally sponsored research in accordance with the DOE Public Access Plan.